\documentclass[fleqn,usenatbib]{mnras}
\usepackage{newtxtext,newtxmath}
\usepackage{txfonts}

\DeclareRobustCommand{\VAN}[3]{#2}
\let\VANthebibliography\thebibliography
\def\thebibliography{\DeclareRobustCommand{\VAN}[3]{##3}\VANthebibliography}

\usepackage{graphicx}	
\usepackage{amsmath}	
\usepackage{amssymb}	
\usepackage{xcolor, color}
\usepackage{booktabs}
\usepackage{threeparttable}
\usepackage{multirow}
\usepackage{comment}
\usepackage{natbib}

\DeclareUnicodeCharacter{2212}{-}
\newcommand{\kms}{km s$^{-1}$}

\newcommand{\meth}{CH$_3$OH}

\title[A train of shocks at 3000 au scale?]{A train of shocks at 3000 au scale?\\ Exploring the clash of an expanding bubble into the NGC 1333 IRAS 4 region. SOLIS XIV}

\author[M. De Simone et al.]{
Marta De Simone,$^{1,2}$\thanks{E-mail: marta.desimone@univ-grenoble-alpes.fr}
Claudio Codella,$^{2,1}$
Cecilia Ceccarelli,$^{1}$
Ana L\'opez-Sepulcre,$^{3,1}$
Roberto Neri,$^{3}$ 
\newauthor Pedro Ruben Rivera-Ortiz,$^{1}$
Gemma Busquet,$^{1}$
Paola Caselli,$^{4}$
Eleonora Bianchi,$^{1,2}$
Francesco Fontani,$^{2,4}$
\newauthor
Bertrand Lefloch,$^{1}$
Yoko Oya,$^{5}$
and Jaime E. Pineda$^{4}$
\\
$^{1}$Univ. Grenoble Alpes, CNRS, IPAG, 38000 Grenoble, France\\
$^{2}$INAF, Osservatorio Astrofisico di Arcetri, Largo E. Fermi 5, 50125 Firenze, Italy\\
$^{3}$Institut de Radioastronomie Millim\'etrique (IRAM), 300 rue de la Piscine, 38400 Saint-Martin d'H\`eres, France\\
$^{4}$Max-Planck-Institut f\"ur extraterrestrische Physik (MPE), Giessenbachstrasse 1, 85748 Garching, Germany\\
$^{5}$Department of Physics, The University of Tokyo, Bunkyo-ku, Tokyo 113-0033, Japan
}

\date{Accepted ---. Received ---; in original form ---}

\pubyear{2021}

\begin{document}
\label{firstpage}
\pagerange{\pageref{firstpage}--\pageref{lastpage}}
\maketitle

\begin{abstract} 
There is evidence that the star formation process is linked to the intricate net of filaments in molecular clouds, which may be also due to gas compression from external triggers.
We studied the southern region of the Perseus NGC 1333 molecular cloud, known to be heavily shaped by similar  {external triggers}, to shed light on the process that perturbed the filament where the Class 0 IRAS4 protostars lie.
We use new IRAM-NOEMA observations of SiO and CH$_3$OH, both known to trace violent events as shocks, toward IRAS 4A as part of the Large Program Seeds Of Life in Space (SOLIS).
We detected three parallel elongated ($>$6000 au) structures, called fingers, with narrow line profiles ($\sim$1.5 km s$^{-1}$) peaked at the cloud systemic velocity, tracing 
gas with high density (5--20$\times10^5$ cm$^{-3}$) and high temperature (80--160 K). 
They are chemically different, with the northern finger traced by both SiO and CH$_3$OH ([CH$_3$OH]/[SiO]$\sim$160--300), while the other two only by SiO ([CH$_3$OH]/[SiO]$\leq 40$).
Among various possibilities, a train of three shocks, distanced by $\geq$5000 yr, would be consistent with the observations if a substantial fraction of silicon, frozen onto the grain mantles, is released by the shocks.
We suggest that the shock train is due to an expanding gas bubble, coming behind NGC 1333 from the southwest and clashing against the filament, where IRAS 4A lies.
Finally, we propose a solution to the two-decades long debate on the nature and origin of the widespread narrow SiO emission observed in the south part of NGC 1333, namely that it is due to unresolved trains of shocks.
\end{abstract}



\section{Introduction} \label{sec:intro}
\begin{table*}
    \centering
    \caption{Spectroscopic parameters and line Gaussian-fit results of CH$_3$OH and SiO lines observed toward the selected positions (1a, 1b and 2) on the IRAS 4A fingers shown in Fig. \ref{fig:ch3oh+sio+hc3n_map}).
    }
    \label{tab:spectral_info+fit}
    \resizebox{\textwidth}{!}{%
    \begin{threeparttable}  
    \begin{tabular}{l|c|ccc|c|cccccc}
    \hline
    \hline
    & \multicolumn{4}{c|}{Spectroscopic parameters} &  & \multicolumn{5}{c}{Gaussian-fit results} \\
    \hline
    Setup$^{(a)}$ & Transition     & Frequency$^{(b)}$	& E$_{\rm up}^{(b)}$  & logA$_{\rm ul}^{(b)}$ & Finger$^{c}$ & Integrated Area  & v$_{\rm peak}$ & FWHM & T$_{\rm peak}$ & rms \\
       &  & [GHz]       & [K]   & [s$^{-1}$]  &  position & [K km s$^{-1}$] & [km s$^{-1}$] & [km s$^{-1}$] & [K] & [mK] \\
    \hline
    & \multicolumn{9}{c}{CH$_3$OH} &\\
    \hline
    \multirow{3}{*}{6-HR} & \multirow{3}{*}{5$_{\rm 1,5}$--4$_{\rm 0,4}$ E}	& 		\multirow{3}{*}{84.5212}	& 	\multirow{3}{*}{40}	&	\multirow{3}{*}{-5.7} & 1a &	25.6(0.3)	& 	6.50(0.01)	&	1.37(0.02) & 17.6 & 300 \\
    & & & & & 1b & 35.3(0.2) & 6.30(0.01) & 1.40(0.01) & 23.7 & 300  \\
    & & & & & 2a & - & - & - & $\leq$0.9$^{d}$ & 300  \\
    \multirow{3}{*}{3-HR} & \multirow{3}{*}{2$_{\rm -1,2}$--1$_{\rm -1,1}$ E} & 		\multirow{3}{*}{96.7394}	& 	\multirow{3}{*}{13}	&	\multirow{3}{*}{-5.6} & 1a &  4.8(0.1) & 6.40(0.02) & 1.60(0.03) & 2.8 & 60 \\
    & & & & & 1b & 8.0(0.1) & 6.30(0.01) & 1.60(0.02) & 4.7 & 60  \\
     & & & & & 2a & - & - & - & $\leq$0.2$^{d}$ & 60  \\
    \multirow{3}{*}{3-HR} & \multirow{3}{*}{2$_{\rm 0,2}$--1$_{\rm 0,1}$ A} 	& 		\multirow{3}{*}{96.7414}	& 	\multirow{3}{*}{7}	&	\multirow{3}{*}{-5.5} & 1a & 4.8(0.1) & 6.20(0.02) & 1.46(0.03) & 3.1 & 60 \\
     & & & & & 1b & 8.8(0.1) & 6.10(0.01) & 1.60(0.02) & 5.2 & 60  \\
      & & & & & 2a & - & - & - & $\leq$0.2$^{d}$ & 60  \\
    \multirow{3}{*}{3-HR} & \multirow{3}{*}{2$_{\rm 0,2}$--1$_{\rm 0,1}$ E} 	&		\multirow{3}{*}{96.7445}	& 	\multirow{3}{*}{21}	&	\multirow{3}{*}{-5.5}  & 1a & 3.5(0.1) & 6.10(0.02) & 1.66(0.05) & 1.9 & 60 	\\
    & & & & & 1b & 4.6(0.1) & 6.00(0.03) & 1.70(0.02) & 2.6 & 60  \\
      & & & & & 2a & - & - & - & $\leq$0.2$^{d}$ & 60  \\
    \multirow{3}{*}{3-HR} & \multirow{3}{*}{2$_{\rm 1,1}$--1$_{\rm 1,0}$ E}  &   	\multirow{3}{*}{96.7555}  	& 	\multirow{3}{*}{28}	&	\multirow{3}{*}{-5.6} & 1a & 1.9(0.1) & 6.40(0.03) & 1.55(0.08) & 1.2 & 60\\  
     & & & & & 1b & 1.8(0.1) & 6.20(0.03) & 1.6(0.1) & 1.1 & 60  \\
      & & & & & 2a & - & - & - & $\leq$0.2$^{d}$ & 60  \\
    \hline
    & \multicolumn{9}{c}{SiO} &\\
    \hline
    \multirow{3}{*}{6-LR} & \multirow{3}{*}{2-1} & \multirow{3}{*}{86.8469} & \multirow{3}{*}{6} & \multirow{3}{*}{-4.5} & 1a & 5.6(0.4) & 7.3(0.5) & 6.9(0.3) & 0.8 & 40\\
     & & & & & 1b & 5.7(0.4) & 8.6(0.4) & 6.9(0.2) & 0.8 & 40  \\
      & & & & & 2a & 6.4(0.3) & 8.3(0.5) & 6.9(0.8) & 0.9 & 40  \\
    
    \hline
    \end{tabular}
    \begin{minipage}{0.5cm}
    \hspace{0.5cm}
    \end{minipage}
    \begin{minipage}{15cm}
    \footnotesize
    $^a$ HR: High resolution ($\Delta$v$\sim$0.5 and 0.2 \kms \ for setup 3 and 6, respectively); LR: Low resolution ($\Delta$v$\sim$6 \kms \ in both setups). \\ 
    $^b$ Frequencies and spectroscopic parameters are taken from \citet{xu_torsion_2008} and \citet{muller_rotational_2013} and retrieved from the CDMS \citep[Cologne Database for Molecular Spectroscopy:][]{muller_cologne_2005} database.
    Upper level energies E$_{up}$ refer to the ground state of each symmetry.
    logA$_{ul}$ are the logarithmic spontaneous emission coefficients.\\
    $^{c}$ Selected position on the IRAS 4A fingers (see text and Fig. \ref{fig:ch3oh+sio+hc3n_map}).\\
    $^{d}$ 3$\sigma$ limit for non-detection.
    \end{minipage}
    \end{threeparttable}
    }
\end{table*}

Molecular clouds forming Solar-type stars are characterized by an intricate net of filaments which are widely accepted to play an important role in the star formation process. 
Indeed most of the young stars and cores in low-mass star forming regions are embedded in filaments of gas and dust which dominate the cloud structure \citep[e.g.,][]{Schneider_catalog_1979,Ungerechts_co_1987,goldsmith_large-scale_2008}, a characteristic shown to be ubiquitous by the large scale maps of the Herschel Space Observatory and Planck satellites \citep[e.g.,][]{molinari_clouds_2010,andre_from_2010, andre_from_2014,zari_herschel-planck_2016}. 
Some of the most prominent cloud filaments actually are collections of velocity-coherent fibers that can become gravitationally unstable and fragment into chains of cores \citep{hacar_cores_2013,hacar_fibers_2017,tafalla_chains_2015,henshaw_seeding_2016,henshaw_unveiling_2017,sokolov_multicomponent_2019,sokolov_probabilistic_2020,chen_velocity-coherent_2020}. 
However, the exact process ruling the formation of these filamentary structures and their specific role in triggering star formation are still debated \citep[e.g.,][]{Hennebelle_role_2019,Robitaille_statistical_2020}. 
A major process in their shaping, in addition to the presence of magnetic fields, is the compression of the molecular gas by external triggers, such as ionization/shock fronts around OB stars or supernovae remnants, cloud-cloud collisions, and (magneto-)hydro-dynamical gravitational instabilities \citep[e.g.,][]{padoan_turbulent_2001,hennebelle_the_2013,Vazquez-Semadeni_global_2019,Federrath_sonic_2021}.

All these external triggers leave signatures at different scales, from parsec to sub-parsec, observed with specific molecular shocks tracers \citep[e.g.,][]{jimenez-serra_parsec-scale_2010,berne_waves_2010,dumas_localized_2014,cosentino_widespread_2018, cosentino_interstellar_2019, cosentino_sio_2020}.
SiO and CH$_3$OH are traditionally considered among the best tracers of such shocks, where their abundance is observed to drastically increase by several orders of magnitude \citep[e.g.,][]{bachiller_molecular_1998,bachiller_chemically_2001,arce_complex_2008,codella_herschel_2012,lefloch_l1157-b1_2017,codella_seeds_2020}.
The SiO enhanced abundance is due to the sputtering of the grain mantles and shattering of the grain refractory cores, both processes releasing SiO and Si (which is quickly oxidized in SiO) into the gas-phase \citep[e.g.,][]{caselli_grain-grain_1997,schilke_sio_1997,gusdorf_sio_2008-1, gusdorf_sio_2008,guillet_shocks_2011}. 
Likewise, \meth \ is believed to be prevalently formed on the cold grain surfaces \citep[e.g.,][]{watanabe_efficient_2002,rimola_combined_2014} and released into the gas-phase by the grain mantle sputtering \citep[e.g.,][]{Flower_methanol_2010}.

Perseus is one of the molecular clouds in the vicinity of our Sun that is believed to have been formed and shaped by external triggers, such as the explosion of one or more supernovae and other forms of stellar feedback activity. 
For example, new 3D dust extinction maps obtained by  {GAIA support this hypothesis} \citep[see e.g.,][]{zucker_three_2021,bialy_per-tau_2021}.
One of the most active sites of ongoing star formation in Perseus is NGC 1333 in the Perseus molecular cloud complex \citep[$\sim$300 pc;][]{zucker_mapping_2018}. 
This region hosts a large number of young stars and protostars associated with filamentary structures, and it displays a complex network of fibers characterized by a high degree of internal fragmentation with typically three to four cores per fiber \citep{hacar_fibers_2017}.

Recent observations produced large scale maps of the magnetic field in NGC 1333, helping in understanding their role in the star formation  {process} \citep[e.g.,][]{doi_jcmt_2020}.
For all these reasons, NGC 1333 is one of the best regions to study the connection between filaments, magnetic fields, external triggers, and star formation.

More specifically, this article focuses on the southern filament of NGC 1333, which is composed of multiple structures with distinct systemic velocities and velocity gradients \citep[e.g.,][]{sandell_ngc_2001,dhabal_morphology_2018,dhabal_connecting_2019}. 
The famous IRAS 4A system, a protostellar binary which drives two large-scale molecular outflows, is located on the west side of the filament \citep{choi_variability_2005, santangelo_jet_2015, de_simone_seeds_2020, taquet_seeds_2020,chuang_alma_2021}. 
The filament joins an arc-like structure on the west which hosts other young protostars (the SVS13 and IRAS 2 systems).
In a recent study, \citet{dhabal_connecting_2019} hypothesized that the entire southwest region of NGC 1333, encompassing the filament where IRAS 4A lie, is due to a colliding ``turbulent cell'', a clash that triggered the birth of the above-mentioned protostars.
However, no specific signatures of a clash, namely shocks, have been reported so far, leaving unanswered how and where the energy of this clash, if real, is dispersed.

To answer this question, we analyzed new high spatial resolution ($\sim 600$ au) observations of \meth \ and SiO toward IRAS 4A, obtained in the context of the IRAM-NOEMA Large Program SOLIS\footnote{\url{https://solis.osug.fr/}} \citep[Seeds Of Life In Space; ][]{ceccarelli_seeds_2017}.
Our goal is to search for specific signatures of the clash event and to gain a more detailed insight into what happened and why.

The article is organized as follows.
In Sects. \ref{sec:obs} and \ref{sec:results}, we described the observations and the results, respectively.
In Sect. \ref{sec:properties}, we analyze the physical and chemical properties of the observed structures.
In Sect. \ref{sec:discussion}, we discuss the implication of our analysis and in Sect. \ref{sec:conclusions} we summarise our conclusions.



\section{Observations} \label{sec:obs}
The IRAS 4A system was observed at 3mm with the IRAM NOEMA (NOrthern Extended Millimeter Array) interferometer\footnote{\url{https://www.iram-institute.org/EN/noema-project.php}} within the SOLIS project$^1$.
Part of the data was already published in \citet{de_simone_seeds_2020}.
For the present study, we used the observations obtained with two setups, hereinafter called 3 and 6, during several tracks (18.6 hrs for setup 3 in June, September, and October 2016 and 6.8 hrs for setup 6 in March 2018).

Setup 3 was observed with the WideX correlator at 95.5-99.5 GHz and with $\sim$2 MHz ($\sim$6 km s$^{-1}$) spectral resolution, plus four narrow spectral windows with $\sim$0.16 MHz ($\sim$0.5 km s$^{-1}$) spectral resolution. 
Setup 6 was observed using the PolyFiX correlator\footnote{\url{https://www.iram.fr/IRAMFR/GILDAS/doc/html/noema-intro-html/node6.html}} at 80.4-88.1 GHz (lower sideband, LSB) and 95.9-103.6 GHz (upper sideband, USB) with $\sim$2 MHz ($\sim$6 km s$^{-1}$) spectral resolution, plus 64 narrow spectral windows with $\sim$0.06 MHz ($\sim$0.2 km s$^{-1}$) spectral resolution.

The NOEMA array was used in configurations DC (baselines 15-304 m) and AC (64-704 m) for setup 3 and 6, respectively.
The phase center was on IRAS 4A1, at coordinates $\alpha({\rm J2000})$ = ${\rm 03^h 29^m 10\farcs5}$, $\delta({\rm J2000})$ = +31$^\circ$ 13$\arcmin$ 30$\farcs$9.

The bandpass was calibrated on 3C454.3 and 3C84, the absolute flux was fixed by observing MWC349, LKHA101, and the gains in phase and amplitude were set on 0333+321. 
The final uncertainty on the absolute flux scale is $\leq$ 15\%.
The phase rms was $\le$ 50$^\circ$ and $\sim$ 10-20$^\circ$ and the typical precipitable water vapor (pwv) was 4-15 mm and 1-30 mm for setup 3 and 6, respectively. 
The system temperature was 50-200 K for both setups.

The data were reduced using the GILDAS\footnote{\url{http://www.iram.fr/IRAMFR/GILDAS}} software collection. 
The continuum map was obtained by averaging line-free channels and self-calibrating the data. 
The self-calibration solutions were then applied to the spectral cubes, which were subsequently cleaned using natural weighting. 
The resulting synthesised beams are 2$\farcs$2$\times$1$\farcs$9 (P.A.=96$^\circ$) and 2$\farcs$05$\times$1$\farcs$12 (P.A.=11$^\circ$) and the half power primary beams are 59$\farcs$2 and 61$\farcs$2 for setups 3 and 6, respectively.

\begin{figure*}
    \centering
    \includegraphics[scale=0.8]{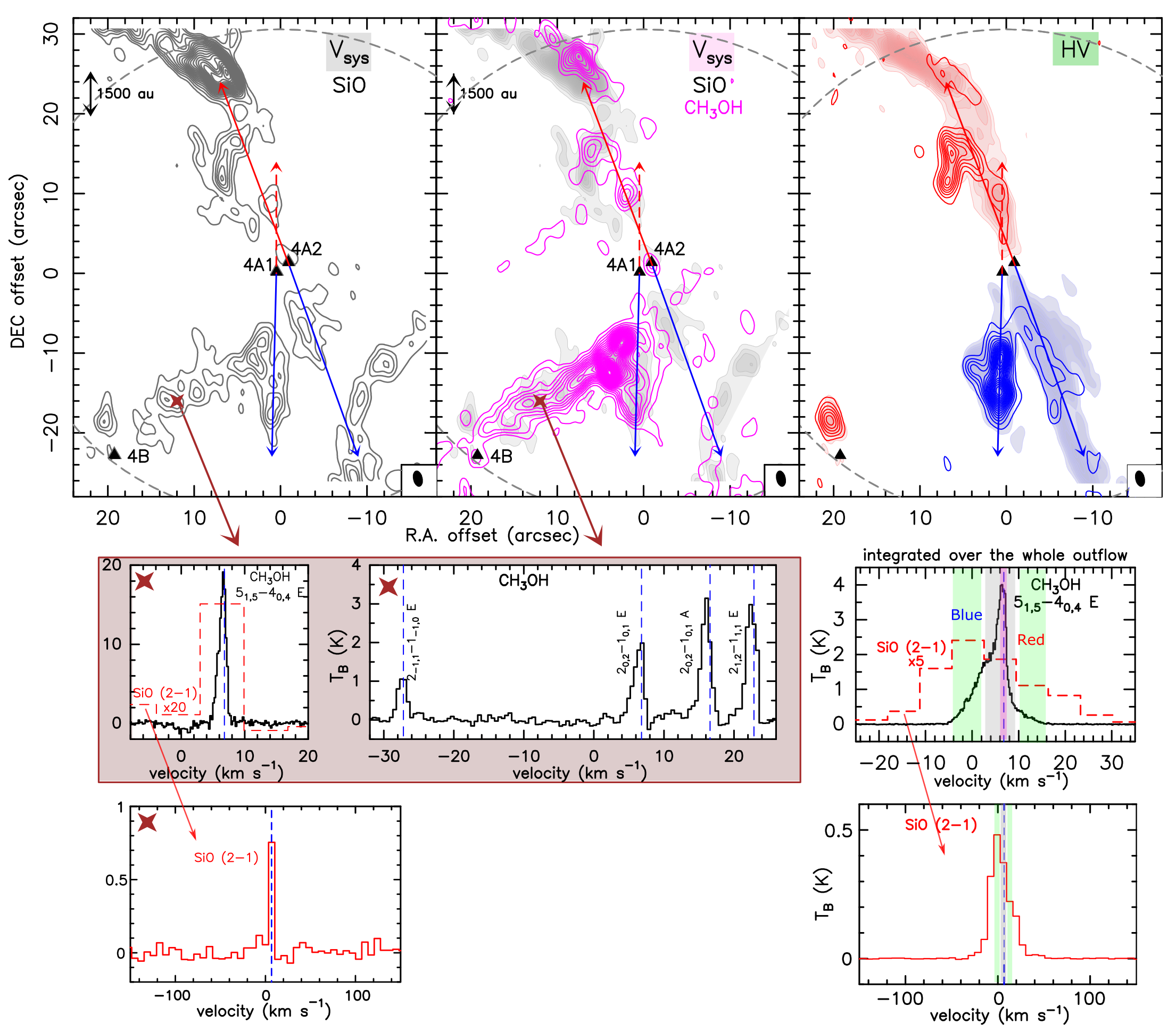}
    \caption{\textit{Upper panels:} Velocity-integrated maps of CH$_3$OH 5$_{1,5}$-4$_{0,4}$ E ({coloured contours}) and SiO 2-1 ({gray contours and }shaded colors) of the NGC 1333 IRAS 4 system. 
    The three protostellar sources in the field are marked by black triangles. 
    Jet directions are indicated by blue and red arrows following \citet{choi_variability_2005} and \citet{santangelo_jet_2015}.
    The primary beam ($\sim 62''$) is shown by the grey dashed circle and the synthesized beam ($\sim 1\farcs5$) is  {shown} in the lower right corner of the panels.
    \ {Upper left panel:} SiO emission at the systemic velocity channel, with first contours and steps of 3$\sigma$ ($\sigma$=1 mJy beam$^{−1}$ \kms).
    \ {Upper middle panel:} CH$_3$OH Emission integrated from 5.8 to 7.8 \kms \ around the systemic velocity, with first contours and steps of 3$\sigma$ ($\sigma$=15 mJy beam$^{−1}$ \kms).
    \ {Upper right panel:} CH$_3$OH Blue- and red-shifted high velocity (HV) emission integrated from -4 to 2.4 \kms \ (blue) and 9.6 to 16 \kms \ (red) with first contours and steps of 3$\sigma$ ($\sigma_{\rm blue}$=35 mJy beam$^{−1}$ \kms \ and $\sigma_{\rm red}$=13 mJy beam$^{−1}$ \kms). 
    {\textit{Lower panels:} CH$_3$OH (in black) and SiO (in red) spectra extracted at the Finger1a position with offset ($-$12$''$, +16$''$) marked by a brown cross (see text). To compare with \meth \ the SiO spectra is overlapped in dashed red magnified by a factor of 20.}
    \ {The map integrated ranges are shown as colored bands on the spectra of SiO (in red) and CH$_3$OH (in black overlapped, for comparison, with SiO in dashed red magnified by a factor of 5) integrated over the whole outflow in the lower right panels}.
    }
    \label{fig:ch3oh_channmaps}
\end{figure*}

\section{Results}\label{sec:results}

\begin{figure*}
    \centering
    \includegraphics[scale=0.7]{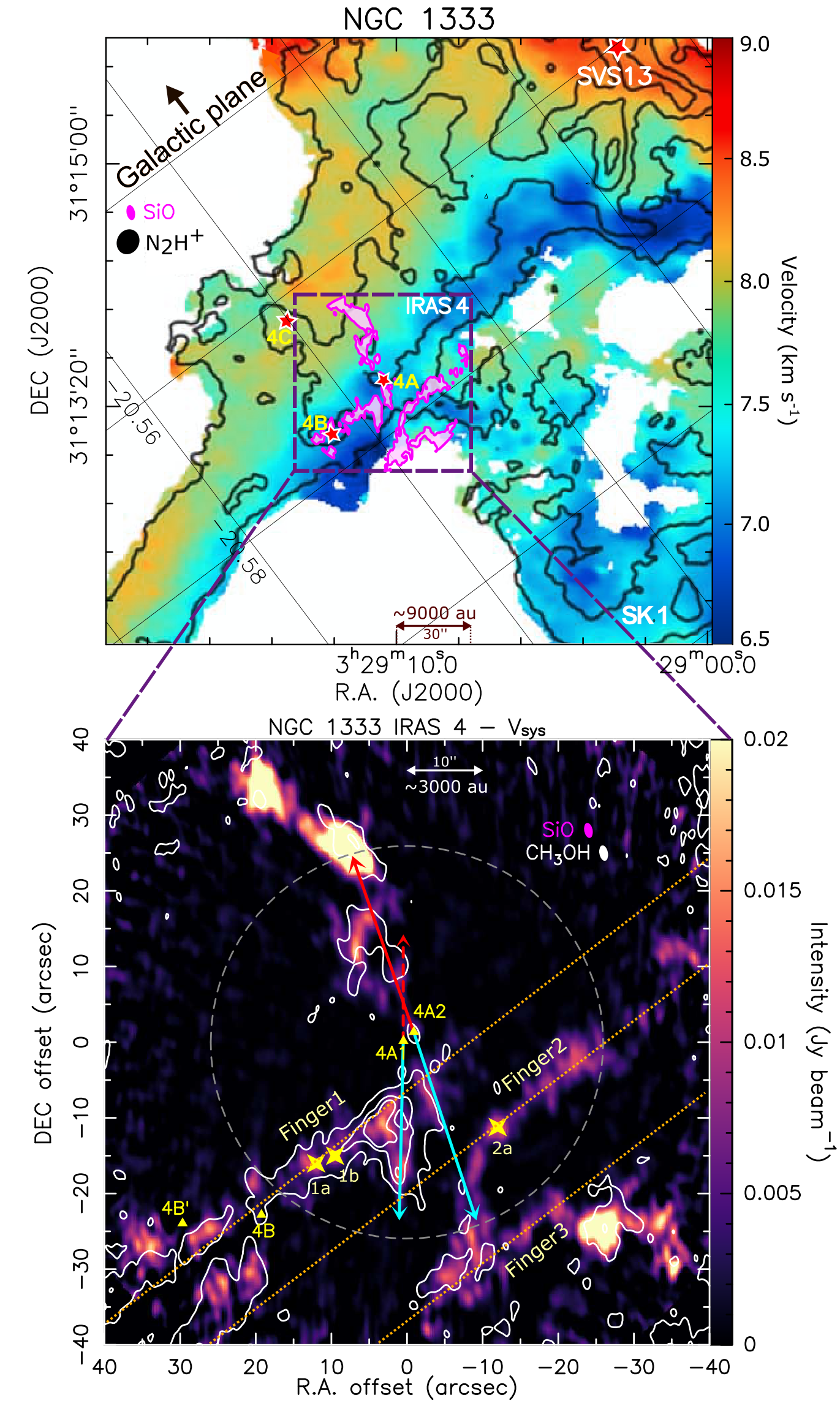}
    \caption{Emission of the fingers detected by NOEMA-SOLIS in the \meth \ and SiO shown at a large scale. 
    \textit{Upper panel}: Overlap of the N$_2$H$^\text{+}$ (1-0) 
    line-of-sight velocity map of the southern-east part of NGC 1333 from CARMA observations \citep[adapted from Fig. 17 of][]{dhabal_connecting_2019}
    with the SiO emission from NOEMA-SOLIS observations (this work)  {shown as 3$\sigma$ contours}. The internal grid is in Galactic coordinates. The red stars mark the position of IRAS 4A, 4B, 4C, and SVS13. The synthesized beams ($\sim 1\farcs5$ for SiO and $\sim 3\farcs5$ for N$_2$H$^\text{+}$) are in the upper left corner. 
    \textit{Bottom panel}: Zoomed-in map of the IRAS 4A system with SiO (colour scale, {3$\sigma$= 3 mJy beam$^{-1}$}) and CH$_3$OH (white contours starting at 3$\sigma$ with steps of 15$\sigma$, $\sigma$=30 mJy beam$^{-1}$) emission integrated in a range of $\sim$6 \kms \, around the v$_{\rm lsr}$ ($\sim$ 6.7 \kms). The yellow triangles indicate the sources 4A1, 4A2, 4B and 4B$'$. 
    The synthesized beams ($\sim 1\farcs5$) are in the upper right corner, while the primary beam ($\sim$61$''$) is a dashed white circle.
    The yellow crosses mark the fingers positions where we carried out a non-LTE analysis of the methanol lines.
    Orange lines show the direction of the three fingers (Finger1, traced by CH$_3$OH and SiO; Finger2 and Finger3 traced by SiO only) and are parallel to the Galactic plane.}\label{fig:ch3oh+sio+hc3n_map}
\end{figure*}

We detected five \meth \ and one SiO lines, listed in Table \ref{tab:spectral_info+fit} with their spectroscopic parameters.
Figure \ref{fig:ch3oh_channmaps} shows the spatial distribution of the CH$_3$OH and SiO line emission in two different velocity ranges: around the systemic velocity V$_{\rm sys}$ ($\sim$ 6.7 \kms) and at High-Velocity (HV), at about $\pm$7 km s$^{-1}$ from the systemic velocity and over a velocity interval of about $\Delta$v$\sim$ 6 \kms \ (see caption for details).
Two major components appear in the figure: the two outflows associated with A1 and A2, especially visible in the HV maps, and two filamentary structures at the systemic velocity, which we will refer to as "fingers" in the following.

\subsection{Outflows emission} 
The emission from the IRAS 4A outflows detected by SOLIS in different molecular lines has been presented already and discussed by \citet{taquet_seeds_2020} (S-bearing species) and \citet{de_simone_seeds_2020} (interstellar Complex Organic Molecules).
The Fig. \ref{fig:ch3oh_channmaps} methanol emission map is in general agreement with those two studies and, being not the focus of this paper, will not be discussed further.

\subsection{Fingers emission} 
The high resolution spectra of the five methanol lines (Fig. \ref{fig:ch3oh_channmaps}) show that the elongated structures are characterized by narrow lines (FWHM$\sim$1.5 \kms) centered at the systemic velocity V$_{\rm sys}$ (see below).
Their filamentary morphology is better highlighted in Fig. \ref{fig:ch3oh+sio+hc3n_map}, where the \meth \ and SiO narrow-line emission is shown over a larger field of view.
Based on Fig. \ref{fig:ch3oh+sio+hc3n_map} and the spectral profile described above, we identified three fingers:\\
\textit{Finger1}, traced by CH$_3$OH and SiO, is the northern one;\\
\textit{Finger2}, traced only by SiO, is 10$''$ south and parallel to Finger1;\\
\textit{Finger3}, traced only by SiO, is an additional 10$''$ south and, again parallel to Finger1.\\ 
While Finger2 and Finger3, detected only in SiO (2-1), are perfectly in agreement with the SiO (1-0) VLA observations by \citet{choi_variability_2005}, Finger1 was never detected before and it is traced by both CH$_3$OH and SiO.
It is worth emphasizing that the three fingers are each separated by about 10$''$ ($\sim$ 3000 au)  {and are parallel to each other}.
Note also that Finger2 and Finger3 seem to be connected to each other by an almost vertical structure, but it is very likely that this connecting structure is tracing the outflowing material from IRAS 4A2.   

In order to understand the nature of the three fingers, we selected three positions along Finger1 and  {Finger2 to carry out} a non-LTE analysis to derive the gas temperature and density and the column density of \meth \ and SiO, respectively ( see Section \ref{subsec:phys-properties})\footnote{Note that since Finger3 is outside the primary beam we did not carry the analysis there.}.
The three positions (two - 1a and 1b - toward Finger1 and one - 2a - toward Finger2), are shown in Fig. \ref{fig:ch3oh+sio+hc3n_map} and their coordinates are reported in Table \ref{tab:LVG_results}.
These positions were selected because they show the brightest \meth \ and SiO emission in the fingers and are outside the outflows emission region in order to minimize the contamination from the latter.
Figure \ref{fig:ch3oh_channmaps} shows the CH$_3$OH and SiO spectra extracted from Finger1a, as an example.

In each of the three selected positions, we extracted the spectra of CH$_3$OH and SiO and derived the velocity-integrated line intensities of each detected transition using a Gaussian fit with the GILDAS-CLASS package. 
The fit results, namely the integrated emission ($\int$T$_b$dV), the linewidth (FWHM), the peak velocities (V$_{\rm peak}$), and the rms computed for each spectral window, are reported in Table \ref{tab:spectral_info+fit}.
As mentioned above, the methanol lines observed with the high resolution narrow bands have an average linewidth of about 1.5 \kms.
Unfortunately, for SiO we are limited by the PolyFiX low spectral resolution ($\sim 6$ \kms). 
However, from the VLA observations, \citet{choi_variability_2005} estimated a similar linewidth ($\sim$1.5 \kms) for the SiO narrow component. 
It is worth noticing that, while the methanol emission drastically changes from one finger to another, the integrated SiO emission derived at points Finger1a, Finger1b, and Finger2a is almost constant.
We will comment the implications of this characteristics in Sect. \ref{sec:properties}.

Finally, we searched for the presence of other molecules in the fingers' positions and only found faint emission of HC$_3$N (11-10), which appears to be associated with Finger1. 

\section{Physical and chemical properties of the Fingers}\label{sec:properties}

\subsection{Physical properties}\label{subsec:phys-properties}
\begin{figure}
    \centering
    \includegraphics[scale=0.6]{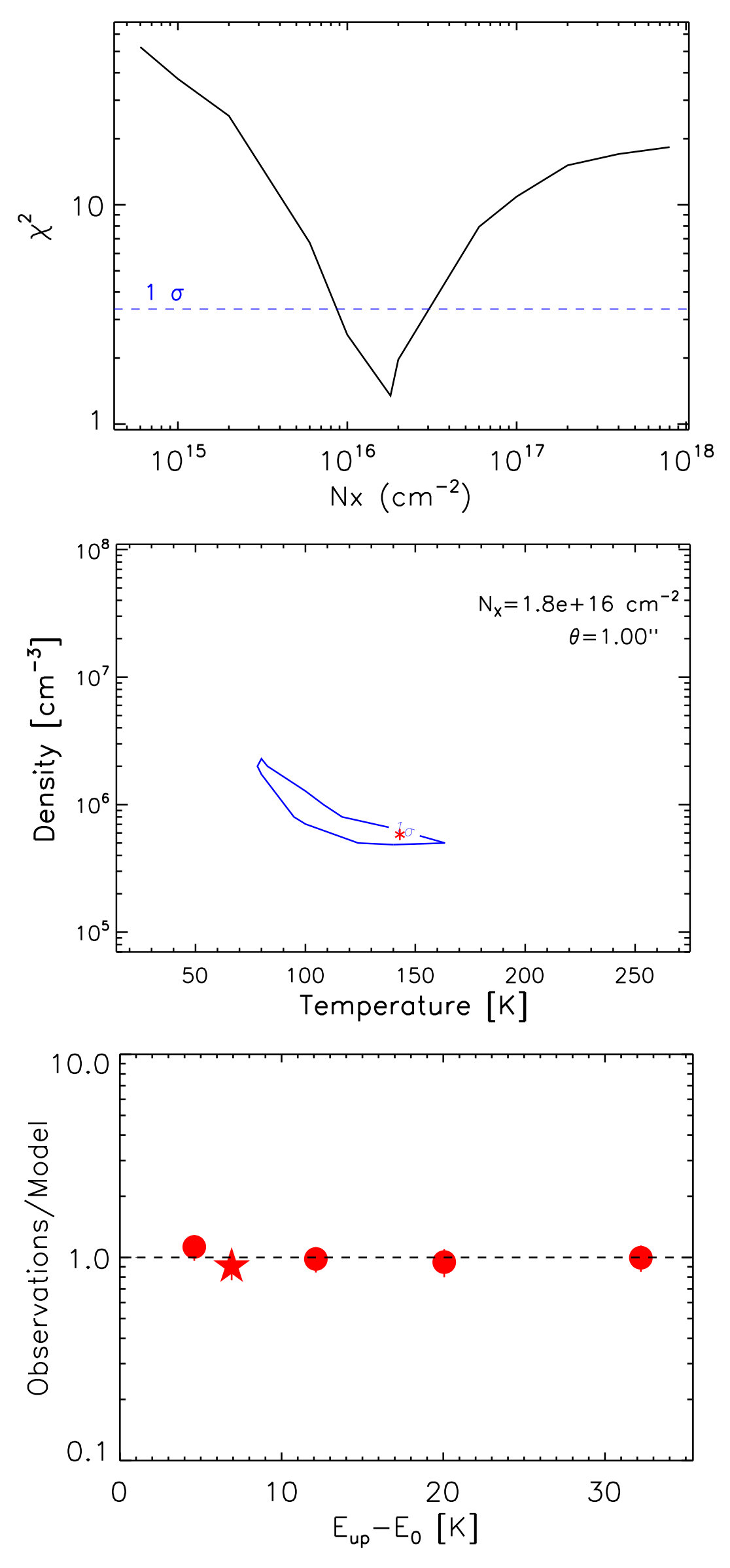}
    \caption{Results of the non-LTE analysis of \meth \  {at} the Finger1a position (see Table \ref{tab:LVG_results}) using the \texttt{grelvg} code. 
    \textit{Top}:  $\chi^2$-column density N(CH$_3$OH) plot. The dashed blue line represents the 1$\sigma$ confidence level.
    \textit{Middle}: Density-Temperature $\chi^2$ contour plot. The contour represents 1$\sigma$ confidence level, assuming the best fit values of N(CH$_3$OH) and $\theta$ (upper right corner). The best fit solution is marked by the red asterisk. 
    \textit{Bottom}: Ratio between the observed line intensities (circles for E-type and stars for A-type) with those predicted by the best fit model as a function of line upper-level energy E$_{\rm up}$.}
    \label{fig:lvg_model}
\end{figure}

\begin{table}
    \centering
    \caption{1$\sigma$ Confidence Level (range) from the Non-LTE LVG Analysis of the CH$_3$OH lines toward the three selected positions in the Fingers (marked as yellow crosses in Fig. \ref{fig:ch3oh+sio+hc3n_map}).}
    \label{tab:LVG_results}
    \hspace{-1cm}
    \resizebox{\columnwidth}{!}{%
    \begin{tabular}{c|cccccc}
         \hline
         \hline
         Species & N(X) & n$_{\rm H_2}$ & T$_{\rm gas}$ & size \\
          & [cm$^{-2}$] & [cm$^{-3}$] & [K] & [arcsec]   \\
          \hline
          & \multicolumn{4}{c}{Finger1a -offset (+12$''$, -16$''$)} \\
          \hline
          CH$_3$OH & 
                     (8-30)$\times$10$^{15}$ & (5-20)$\times$10$^5$ & 80-160 & 0.6-1.5 \\
          SiO & (5-10)$\times$10$^{13}$ & $''$ & $''$ & $''$  \\
          \hline
          [CH$_3$OH]/[SiO] & 160-300 & \\
          \hline
          & \multicolumn{4}{c}{Finger1b - offset (+9.5$''$, -15$''$)} \\
          \hline
          CH$_3$OH & 
                     (4-12)$\times$10$^{15}$ & (2-3)$\times$10$^5$ & 130-210 & 1.5-2.5 \\ 
          SiO & (2-5)$\times$10$^{13}$ & $''$ & $''$ & $''$  \\
          \hline
          [CH$_3$OH]/[SiO] & 200-240 & \\
          \hline
          & \multicolumn{4}{c}{Finger2a - offset (-12$''$, -11.2$''$)} \\
          \hline
          CH$_3$OH$^a$ & 
                        $\leq$1.6$\times$10$^{15}$ & (5-20)$\times$10$^5$ $^a$ & 80-210 $^a$ & 1-2 $^a$ \\
          SiO & (4-15)$\times$10$^{13}$ & $''$ & $''$ & $''$  \\
          \hline
          [CH$_3$OH]/[SiO] & $\leq$40 & \\
          \hline
          \multicolumn{5}{l}{\footnotesize $^a$ Assumed the same gas condition (n$_{\rm H_2}$ and T$_{\rm gas}$) of the Finger1.}
    \end{tabular}
    }
    
\end{table}

The detection of five CH$_3$OH transitions allowed us to perform a non-LTE analysis via our in-home Large Velocity Gradient (LVG) code \texttt{grelvg} adapted from \citet[][]{ceccarelli_theoretical_2003}.

CH$_3$OH has two nuclear-spin isomers, A-type (symmetric) and E-type (asymmetric), differentiated by the total spin state of the hydrogen nuclei in the CH$_3$ group \citep{rabli_rotational_2010}.
We used the collisional coefficients of both CH$_3$OH isomers with para-H$_2$, computed by \citet{rabli_rotational_2010} between 10 and 200 K for the first 256 levels and provided by the BASECOL database \citep{dubernet_basecol2012:_2013}.
We assumed the CH$_3$OH-A/CH$_3$OH-E ratio equal to 1.
A semi-infinite slab geometry was used to compute the line escape probability as a function of the line optical depth.

We ran a large grid of models ($\geq 5000$) covering a total methanol (A-type plus E-type) column density N(CH$_3$OH) from 6$\times$10$^{14}$ to 8$\times$10$^{17}$ cm$^{−2}$, a gas H$_2$ density n$_{\rm H_2}$ from 3$\times$10$^{5}$ to 6$\times$10$^{6}$ cm$^{−3}$, sampled in logaritmic scale, and a gas temperature T from 20 to 200 K.
We then simultaneously fitted the measured CH$_3$OH-A and CH$_3$OH-E line intensities by comparing them with those predicted by the \texttt{grelvg} model, leaving N(CH$_3$OH), n$_{\rm H_2}$, T, and the emitting size $\theta$ as free parameters.
Following the observations, we assumed the linewidth equal to 1.5 \kms \ (see Table \ref{tab:spectral_info+fit}), and we included the flux calibration uncertainty (15\%) to the observed intensities errors.

The best fit in the Finger1a position is obtained for a total column density N(CH$_3$OH)=1.8$\times$10$^{16}$ cm$^{−2}$, gas density n$_{\rm H_2}$=5$\times$10$^5$ cm$^{-3}$, gas temperature T=140 K  and emitting size of $\sim 1''$ with reduced $\chi^2_R$=1.3 (see Figure \ref{fig:lvg_model}). 
The values within the 1$\sigma$ confidence level are reported in Table \ref{tab:LVG_results}. 
To explore the gas conditions along the Finger1, we repeated the analysis described above in position 1b still finding high gas temperature and density (see Table \ref{tab:LVG_results}). 
The derived emitting sizes represent the best-fit 2D-Gaussian FWHM\footnote{In fitting the observations with the LVG theoretical predictions, we considered three possibilities for the filling factor, depending on the shape of the emitting region: i) a 2D circular Gaussian shape, ii) an infinite finger-like shape with a resolved transverse size, and iii) emitting size larger than the synthesised beam.  {The best fit by far} was obtained with the 2D circular Gaussian shape.}, and suggest that the fingers may have a clumpy structure, not resolved by our $\sim$600 au spatial resolution.

To derive the SiO column density in the three positions we run \texttt{grelvg} for a range of N(SiO) from 5$\times$10$^{12}$ to 1$\times$10$^{15}$ cm$^{-2}$ assuming the same gas temperature, density and emitting size as those found in Finger1a (Table \ref{tab:LVG_results}). 
We then best fitted the measured SiO line intensities via comparison with those simulated by \texttt{grelvg} leaving only N(SiO) as a free parameter.
The same procedure was adopted to estimate the upper limit of the CH$_3$OH column density in Finger2a.
The results of the non-LTE analysis carried out toward the three positions are summarised in Table \ref{tab:LVG_results}. 

In order to assess the impact of assuming that the gas conditions derived for \meth \ are also valid for SiO, we estimated the SiO column density by combining our NOEMA observations with the VLA observation by \citet{choi_variability_2005}. 
We performed a LTE analysis using the SiO 2-1 and SiO 1-0 integrated emission in Finger2a deriving a SiO column density of ($30 \pm 7) \times 10^{12}$ cm$^{-2}$, which is consistent with the SiO column density derived above following our assumption (see Table \ref{tab:LVG_results}).
In other words, even if the gas conditions in Finger1 and Finger2 are not exactly the same, the observed abundance ratio of the two species is consistent, within the error bars, with what we computed.
Additionally, in order to investigate the non detection of SiO in Finger1 in the maps of \citet{choi_variability_2005}, we computed the predicted intensity ratio of the SiO 2-1 (target of our NOEMA observations) with respect to the SiO 1-0 \citep[target of the VLA observations of][]{choi_variability_2005} using the non-LTE LVG \texttt{grelvg} code. 
We found that the predicted SiO 2-1/1-0 line ratio varies from 6.5 to 13 in the range of gas density and temperature derived in Finger1 with our analysis (Table \ref{tab:LVG_results}). 
Considering the various uncertainties, in particular on the column density in Finger1 and Finger2 (which are assumed to be the same but can be also a factor $\sim$2 different), there is no contradiction with the fact that there is no SiO detection in Finger1 with the VLA observations by \citet{choi_variability_2005}.


\subsection{Chemical properties}\label{subsec:chem_prop}

From the computations of the previous subsection, we can estimate the \meth \ and SiO abundances in Finger1 as follows.
In Finger1, assuming that the finger depth is equal to the linear diameter of the methanol emitting region ($\sim0.6-1.5''$ equivalent to $\sim180-450$ au: see Tab. \ref{tab:LVG_results}) and considering the derived gas density range ($\sim 5-20\times10^5$ cm$^{-3}$: see Tab. \ref{tab:LVG_results}), we obtained the H$_2$ column density N(H$_2$) range equal to $\sim1.4-14\times10^{21}$ cm$^{-2}$.
Using the derived CH$_3$OH column density range ($\sim 8-30\times10^{15}$ cm$^{-2}$: Tab. \ref{tab:LVG_results}), we then estimate a methanol abundance [CH$_3$OH]/[H$_2$] range of $0.6-22\times 10^{-6}$.
Likewise, the SiO abundance [SiO]/[H$_2$] range is estimated to be 0.4--7.4 $\times$ 10$^{-8}$. 
The SiO abundance is then lower than that measured in the IRAS4A jets, 3--4 $\times$ 10$^{-7}$ by \citet{santangelo_jet_2015}.

From the derivation of the \meth \ and SiO column  {densities} in Finger1 and Finger2 of the previous subsection, the [CH$_3$OH]/[SiO] ratio is 160--300 in Finger1 and $\leq$40 in Finger2 (see Tab. \ref{tab:LVG_results}). 
There seems to be a difference in the chemical composition of Finger1 and Finger2, at least regarding \meth \ and SiO, with Finger1 enriched in methanol with respect to Finger2.

\section{Discussion}\label{sec:discussion}
\begin{figure*}
    \centering
    \includegraphics[scale=0.7]{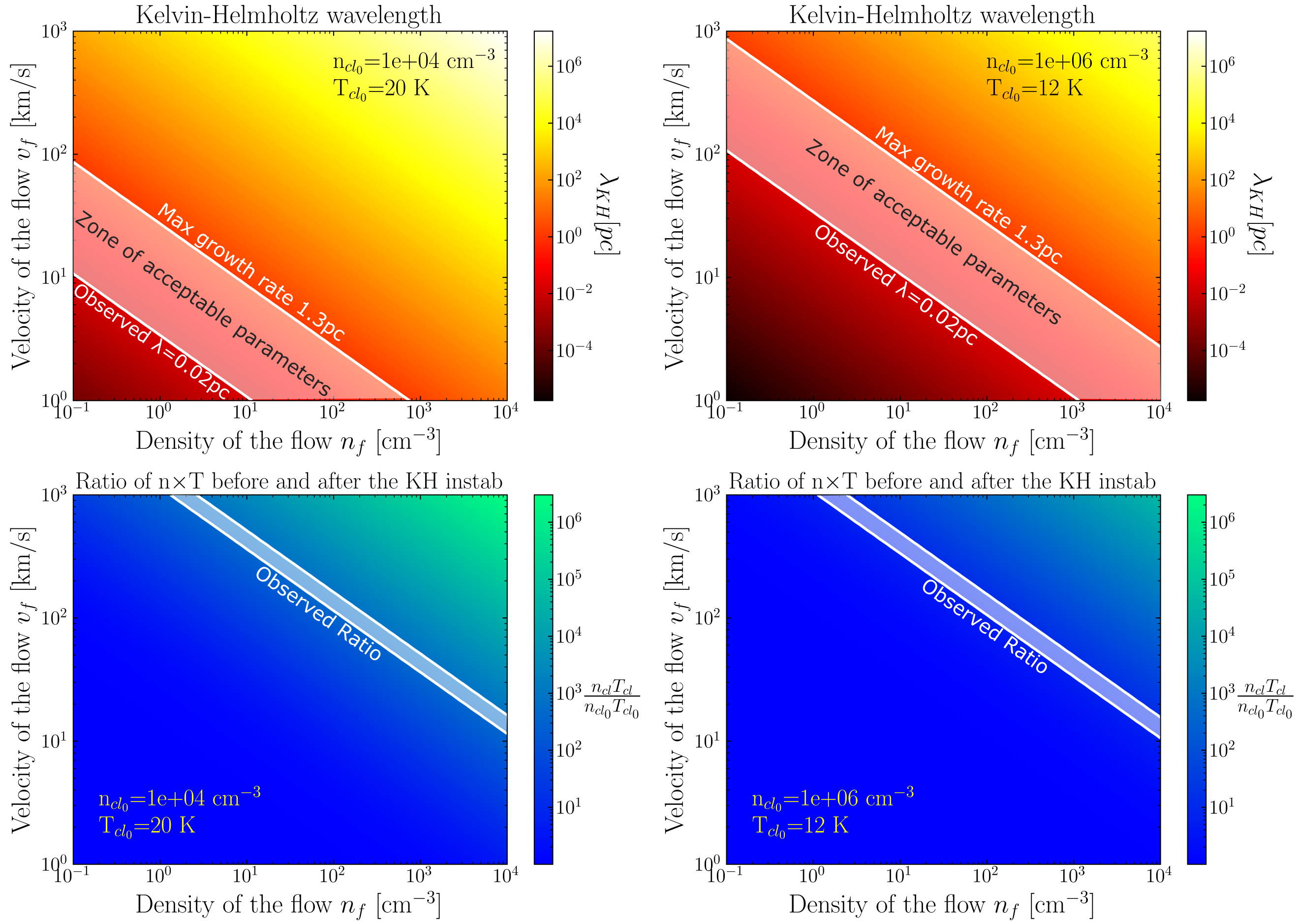}
    \caption{\textit{Top:} Evolution of the Kelvin-Helmholtz wavelength for the observed molecular cloud as a function of density (n$_f$) and velocity of the flowing gas (v$_{f}$). The white continuous lines represent the observational constraints that set the zone (shadowed white) of physically acceptable values for n$_f$ and v$_f$ that allow a Kelvin-Helmholtz instability to exist. The condition assumed for the cloud before the interaction are reported in the upper right corner.
    \textit{Bottom:} Ratio of the density temperature product before and after the KH instability as a function of density (n$_f$) and velocity of the flowing gas (v$_{f}$). The shadowed white zone represents the observed ratio. The conditions assumed for the cloud before the interaction are  {given} in the lower left corner.}
    \label{fig:KH_plots}
\end{figure*}
\subsection{Summary of the fingers properties}\label{subsec:discussion-properties-summary}
In the previous sections, we have shown that the new SOLIS maps, coupled with older VLA maps, reveal the presence of three filamentary structures, extended more than about 6000 au in the major axis (i.e. $\geq$ 20$''$) and unresolved in the minor one, namely $\leq$450 au (Fig. \ref{fig:ch3oh+sio+hc3n_map} and Tab. \ref{tab:LVG_results}).
These structures, which we named "fingers", are approximately aligned in the direction of the galactic plane and to the large scale magnetic field observed by Planck \citep{PlanckCollaboration2020a}, but  {approximately} perpendicular to the local magnetic field \citep[e.g.][; see their Fig. 4]{doi_jcmt_2020} and the SE filament where IRAS 4A lies.
The three fingers are regularly spaced by about $\sim 3000$ au (i.e. $\sim$10$''$).
Finally, they are almost perpendicular (certainly not parallel) to the outflows emanating from IRAS 4A1 and 4A2, even though they seem connected with them in some way: Finger1 starts midway at the edge of the south lobe of the 4A1 outflow and extends east of it, Finger2 starts midway at the edge of the south lobe of the 4A2 outflow and extends west of it, and Finger3 starts at the visible south end of the 4A2 outflow and extends west of it.

Using the detected multiple methanol lines, we carried out a non-LTE analysis that provided stringent constraints to the temperature and density of the gas in Finger1 (Section \ref{subsec:phys-properties}; Tab. \ref{tab:LVG_results}).
The derived temperature, 80--160 K, is definitively larger than that in the surrounding envelope gas, estimated to be smaller than $\sim$12 K \citep{jorgensen_physical_2002,maret_ch3oh_2005}.
In other words, Finger1 is too far away from the IRAS 4A for its gas to be heated by the protostar radiation, so that a non-thermal process must be responsible for the observed large temperature.

In the same vein, the measured SiO enhanced abundance $4-7.4 \times 10^{-8}$ (Section \ref{subsec:chem_prop}) points to a non-thermal process capable of extracting silicon from the grains (see Introduction and Section \ref{subsec:discussion-train}).
Likewise, methanol in Finger1, with an abundance of $0.6-22 \times 10^{-6}$,  {suggests} a similar non-thermal process capable to extract it from the grain mantles.

There are not many possible non-thermal processes capable to create the observed linear and almost periodic structures with enhanced gas temperature and capable to increase the SiO and methanol abundances.
To our best knowledge, only two processes could a priori reproduce the observed properties of the fingers in NGC 1333 IRAS 4A: hydro-dynamical Kelvin-Helmholtz instabilities and shocks.
In the following, we will discuss in detail these two possibilities.

\subsection{Do the IRAS 4A fingers trace Kelvin-Helmholtz instabilities?}\label{subsec:discussion-hdi} 
Hydro-dynamical instabilities (HDI) are known to play a major role in shaping the morphology of the ISM, in particular in modifying the superficial structure of molecular clouds via Rayleigh-Taylor, Kelvin-Helmholtz, and self-gravity instabilities. 
All these processes can generate long and narrow streams of material \citep{hunter_kelvin-helmholts_1997, coughlin_gravitational_2020}.
Among them, the Kelvin-Helmholtz instability (KHI) occurs at the interface of two fluids of different densities in relative shear motion and it is characterized by a wavelike periodic structure. 
Such instabilities have been invoked to explain the periodic filamentary structures observed in  Orion \citep{berne_waves_2010, berne_kelvin-helmholtz_2012} and Taurus \citep{heyer_striations_2016}.
Following the approach by \citet{berne_waves_2010}, we explored if the development of a KHI is possible and in what conditions, and if it can reproduce the gas properties of the IRAS 4A fingers, assuming that the insulating layer (i.e., the interface between the two fluids) is infinitely thin.

\subsubsection{Method} 
First, the maximum spatial wavelength $\lambda_{KH}$ of the KHI is connected to the physical conditions in which the instability occurs. 
Specifically, the value of $\lambda_{KH}$ depends  {on the acceleration due to} self-gravity $g_{cl}$ and on the density $n_{cl}$ of the first cloud, and the relative velocity v$_{f}$ and density $n_{f}$ of the second cloud, which we will call ``flowing cloud'', as follows:  
\begin{equation}\label{eq:KH_wavelenght}
    \lambda_{KH} \  \text{=} \ \frac{2\pi}{g_{cl}} {\rm v}_f^2\frac{n_f}{n_{cl}}
\end{equation}
where we used $g_{cl}$=$\pi$ G$\mu$ m$_H$ N$_H$ and which gives $1\times10^{-11}$m s$^{-2}$ assuming N$_H$ equal to 1.2$\times$10$^{21}$ cm$^{−2}$ (from our analysis, see Section \ref{subsec:chem_prop}).
On the other hand, the growth rate $\omega_{KH}$ for a Kelvin Helmholtz instability depends on the velocity v$_{f}$ and density $n_{f}$ of the flowing cloud as follows: 
\begin{equation}\label{eq:KH_growth_rate}
    \omega_{KH}^2 \ \text{=} \ \frac{k^2 {\rm v}_f^2 n_f n_{cl}}{\text{(} n_f \ \text{+}\ n_{cl}\text{)}^2} \ ,
\end{equation}
where $k$ is the spatial wavenumber. 
If the KHI is responsible for the IRAS 4A fingers, their spatial separation sets a lower limit to $\lambda_{KH}$ equal to $\sim$0.02 pc at the distance of NGC 1333 (and assuming a face-on orientation).
Likewise,  the widths of the lines that trace the fingers set an upper limit to $\omega_{KH}$ of about $8\times10^5$ yrs$^{-1}$.  

Second, we estimated how the temperature and density of the cloud would change in the region of KHI growth.
As the first assumption, we considered the two interacting fluids in hydrostatic equilibrium so that the pressure of the flow and the cloud before the interaction is the same ($P_{f_0}$=$P_{cl_0}$). 
This is a standard assumption since the fluids previously were not mixed. 
When the instability starts to grow, the thermal pressure and the ram pressure of the flow interact with the instability so that:
\begin{equation}\label{eq:pressure-balance}
P_{cl}\  \text{=} \ P_{cl_0} \ \text{+} \ \mu m_H n_f\text{(}{\rm v}_f \cos\theta\text{)}^2 \ ,
\end{equation}
where $\theta$ is the angle between the flow and the instability. 
Assuming that the vertical amplitude is unlikely to be larger than $\lambda$, the angle between the instability interface and the flow is at most $\theta$=45$^\circ$, and considering the ideal equation of state ($P$=$nk_BT$, where $k_B$ is the Boltzmann’s constant) we can rewrite Eq. \ref{eq:pressure-balance} to link the density and temperature before ($n_{cl_0}$ and $T_{cl_0}$) and after ($n_{cl}$ and $T_{cl}$) the KHI emergence, respectively, as follows:
\begin{equation}\label{eq:nT_ratio}
    \frac{n_{cl}T_{cl}}{n_{cl_0}T_{cl_0}} \  \text{=} \ \frac{m_H n_f {\rm v}_f^2}{2n_{cl_0} k_B T_{cl_0}} \ \text{+} \ 1 \ ,
\end{equation}
We considered two cases: 
i) the interaction occurs in the cloud, where the gas density is $10^4 $ cm$^{-3}$ and the temperature is 20 K \citep[i.e. the same assumptions of][]{berne_waves_2010}; 
ii) the interaction takes place inside the IRAS 4A envelope, at $\geq$3000 au from the center (where the first finger is located in the 2D projection), where the gas temperature and density are estimated to be $\leq 12$ K and $\leq10^6$ cm$^{-3}$ \citep[][]{jorgensen_physical_2002,maret_ch3oh_2005}. 

\subsubsection{Results} 
Figure \ref{fig:KH_plots} shows the results of our calculations.
Specifically, it displays the evolution of the KHI wavelength (Eq. \ref{eq:KH_wavelenght}) and the density-temperature ratio before and after the KHI (Eq. \ref{eq:nT_ratio}), using a grid of possible values for the velocity and density of the flowing cloud (v$_f$ from 1 to 10$^3$ \kms and $n_f$ from 0.1 to 10$^4$ cm$^{-3}$), and for the two cases described above (KHI occurring in the cloud or IRAS 4A envelope). 
Whatever are the conditions of the flowing gas, Fig. \ref{fig:KH_plots} demonstrates that KHI can not explain at the same time the observed fingers separation and the observed local increase of density and temperature in the cloud. 

Finally, in addition to the difficulty to reproduce the observed fingers separation and enhanced gas density and temperature, KHI can not explain the presence of \meth \ and SiO in the gas.
Since no sputtering or shattering are expected to play a major role in KHI, the only way for them to release \meth \ and SiO from the grains is via thermal evaporation of their volatile mantles.
However, although the gas cools down slowly via line emission, the dust would cool down very rapidly and probably would never reach the temperature necessary for the grain mantle to sublimate.

\textit{In summary}, KHI does not seem able to explain the observed properties of the IRAS 4A fingers.

\subsection{Do the IRAS 4A fingers trace a train of shocks?}\label{subsec:discussion-train}
\begin{figure}
    \centering
    \includegraphics[scale=0.5]{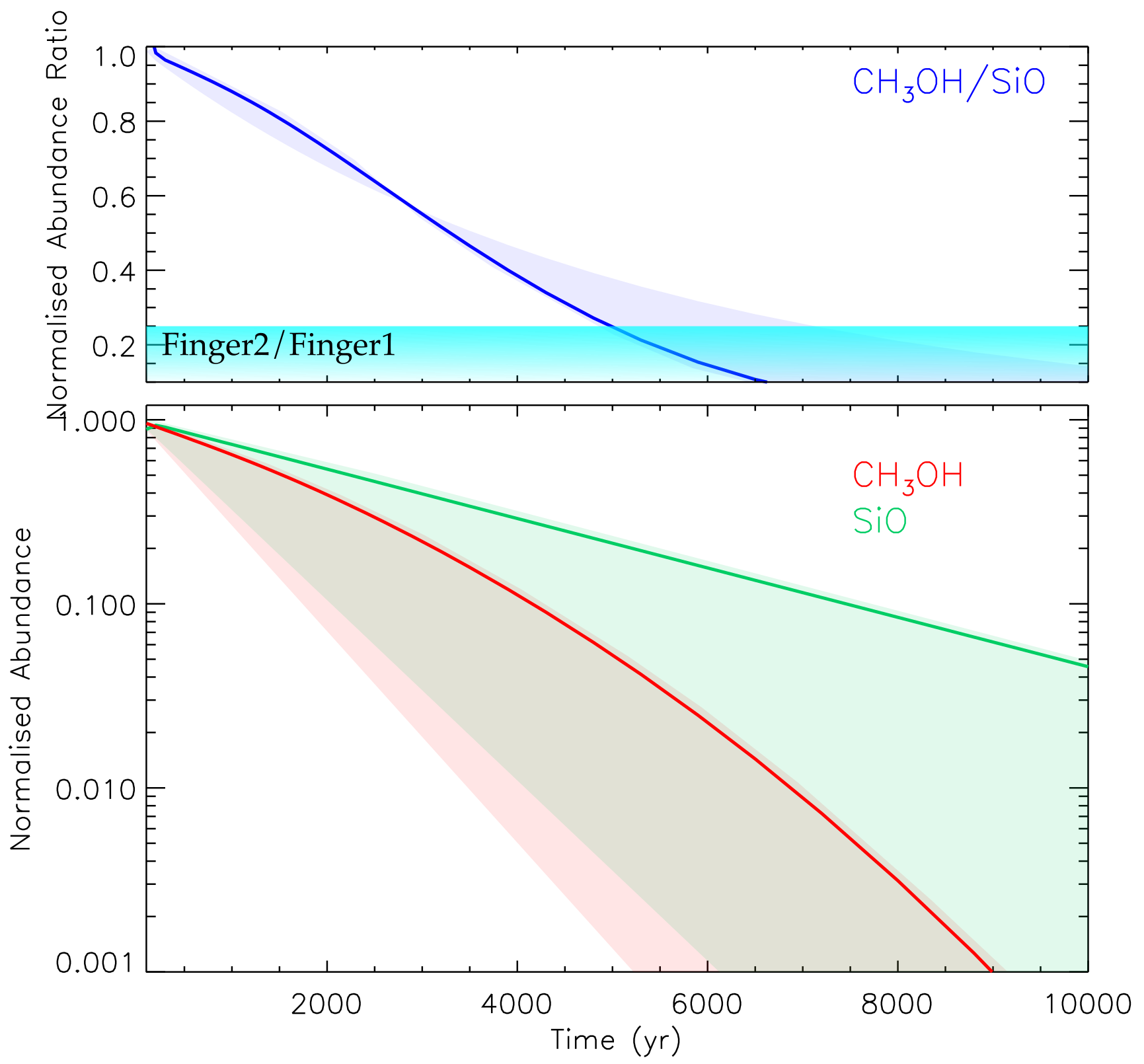}
    \caption{Predicted chemical evolution of methanol (red lines) and SiO (green lines) once injected into the gas phase by the passage of a shock. 
    The solid lines show the predictions using the non-LTE best fit (Section \ref{subsec:phys-properties}), namely $\sim 140$ K and $\sim 5\times10^5$ cm$^{-3}$.
    The shadowed area represents the range of predictions obtained using the limits on the gas density and temperature derived by the non-LTE analysis and listed in Table \ref{tab:LVG_results}.
    \textit{Bottom panel:} Abundances normalized to 1 as a function of time.  
    \textit{Top panel:} Normalised [CH$_3$OH]/[SiO] abundance ratio as a function of time. 
    The horizontal band represents the range of normalized abundance ratio observed in Finger2 with respect to that observed in Finger1.}
    \label{fig:chem_model}
\end{figure}
Shocks are omnipresent in the ISM, at various scales and in various objects, playing a major role in its thermal and physical structure.
In particular, for decades young forming stars are known to be the source of energetic ejections of material which causes shocks when they encounter the quiescent surrounding matter \citep[e.g.,][]{lada_cold_1985}.
In these shocks, the gas and density temperature, as well as both SiO and methanol abundances are enhanced \citep[e.g.,][]{bachiller_molecular_1998,bachiller_chemically_2001,arce_complex_2008,codella_herschel_2012,lefloch_l1157-b1_2017,codella_seeds_2020}.
The quantitative details depend on the type of shock, jump (J) or continuous (C) when it is in presence of a magnetic field, the pre-shocked gas properties and velocity of the shock \citep[e.g.,][]{Hollenbach1979, draine1993}.
In the following, we discuss whether the observed properties of the fingers can be accounted for by shocks. 

\subsubsection{The shock hypothesis}
The analysis of the relative abundances of \meth \ and SiO shows a clear difference in the chemical composition of Finger1 and Finger2, with Finger1 enriched in methanol with respect to Finger2 by more than a factor four. 
As mentioned in the Introduction, methanol can not be produced by gas-phase reactions in the measured quantity \citep{geppert_dissociative_2006}, while it is easily formed on the grain mantle surfaces by the hydrogenation of frozen CO \citep[e.g.,][]{tielens_model_1982,watanabe_efficient_2002, rimola_combined_2014} during the prestellar cold phase \citep[e.g.,][]{caselli_our_2012}.
Therefore, the gaseous methanol observed in Finger1 must have been extracted from the grain mantles and injected into the gas phase.
A similar argument applies to the observed gaseous SiO.
 {The abundance of Si} is extremely low in molecular clouds \citep[$\leq 10^{-12}$;][]{ziurys_shock_1989,requena-torres_organic_2007} because it is trapped in the refractory and, to a lesser extent, in the volatile components of interstellar grains \citep{jenkins_unified_2009}, but it becomes very abundant in the gas where it is extracted from the grains, such as in shocks.
In summary, the gaseous SiO and CH$_3$OH observed toward the three fingers must originate by their injection from the interstellar grains.
This implies two possible explanations of the observed different [CH$_3$OH]/[SiO] abundance ratio: either (1) the two species are injected from the grains with a different abundance ratio in the different fingers or (2) they are injected from the grains with the same abundance ratio but the latter changes in time because of chemical processes occurring in the gas phase.
In the following, we investigate whether the relative and absolute abundances can be reproduced by shocks with properties appropriate for the IRAS 4A fingers (the details of these properties will be described in the following discussion). 

\subsubsection{Chemical modeling of the measured [CH$_3$OH]/[SiO] abundance ratio: time constraints on the shock passage}\label{subsec:time-constraint}
To quantify the SiO and methanol abundance evolution after their injection into the gas phase and whether it could explain the measured [CH$_3$OH]/[SiO] ratio, we ran an astrochemical model and compared the predictions with the observations in the three fingers.
To this end, we used the time-dependent gas-phase code MyNahoon, which is a modified version of the publicly available code Nahoon\footnote{\url{http://kida.astrophy.u-bordeaux.fr/codes.html} } \citep{wakelam_kinetic_2012}, in which we added the accretion of gaseous species into the grain mantles (but no chemistry is computed on the grain surfaces).
To describe the injection of SiO and CH$_3$OH in the gas-phase, we adopted a two-step procedure, as follows \citep[see also][]{codella_seeds_2017,codella_seeds_2020,de_simone_seeds_2020}.

\noindent
\textit{Step 1:} We first compute the chemical composition of the gas during the phase preceding the ejection phenomenon. 
For this, we assumed the steady-state composition of a molecular cloud at 10 K with a H$_2$ density of $2 \times 10^4$ cm$^{−3}$, cosmic-ray ionization rate of $3\times10^{-17}$ s$^{-1}$, visual extinction of 20 mag, and initial elemental abundances as those adopted by \citet[][ their Table 3]{agundez_chemistry_2013}.

\noindent
\textit{Step 2:}  We increase the gas temperature, density, and gaseous abundance of the major components of the grain mantle to simulate their injection into the gas-phase due to the passage of the shock, while the other species abundance results from Step 1. 
The gas temperature and H$_2$ density are those found by the non-LTE LVG modeling in the Finger1a (Table \ref{tab:LVG_results}). 
The abundances of the species injected into the gas phase are assumed to be those measured by IR observations of the interstellar dust ices in similar regions \citep{boogert_observations_2015}:
$2\times10^{−4}$ for H$_2$O; $3\times10^{−5}$ for CO$_2$, CO and CH$_4$;  $2\times10^{−5}$ for CH$_3$OH and NH$_3$, where the abundances are with respect to the H-nuclei.
The SiO  {abundance} is assumed to be 200 times smaller than CH$_3$OH, namely $1\times10^{−7}$.

\noindent
\textit{Results:} 
The evolution of the SiO and CH$_3$OH abundances and their ratio are shown in Fig. \ref{fig:chem_model}. 
Using the best fit of the non-LTE analysis (Section \ref{subsec:phys-properties} and Table \ref{tab:LVG_results}), namely $\sim 140$ K and $\sim 5\times10^5$ cm$^{-3}$, once methanol is ejected into the gas phase, its abundance decreases by a factor of 10 after 4000 yr mainly due to the reaction with OH \citep{shannon2013}, while SiO remains in the gas phase longer (decreasing by a factor of 10 in abundance around 7000 yr because of the freeze-out into the grains).
This basic result holds also for relatively different abundances of SiO and CH$_3$OH, with a ten times abundance decrease of methanol in 2000 yr or more and of SiO in 3000 yr or more.
More important in the context of understanding what happened, for their normalized ratio being as that observed in Finger1 and Finger2 the difference in age of the two shocks should be larger than about 5000 yr (Fig. \ref{fig:chem_model}, top panel).
In other words, if the same relative quantity of \meth \ and SiO has been injected into the gas phase by the passage of two shocks, at Finger1 and Finger2 respectively, the shock in Finger1 is younger more than 5000 yr compared to that in Finger2.
However, if the relative initial abundance ratio is different by a factor two, then the the time between two shocks can be different, namely 2000 yr if Finger2 has an initial abundance ratio higher than Finger1 and $10^4$ yr in the opposite case.

\subsubsection{Shock modeling of the SiO injection: constraints on the shock velocity}\label{subsec:svel-constraint}
Since more than two decades models have predicted that the SiO abundance is highly enhanced in high velocity ($\geq 25$ km/s) shocks, where the grains are shattered and/or sputtered and silicon is liberated into the gas-phase where it undergoes reactions leading to the SiO formation \cite[e.g.][]{draine1983,flower1994,caselli_grain-grain_1997,schilke_sio_1997}.
In lower velocity shocks, Si or SiO previously frozen onto the grain mantles are predicted to be injected into the gas-phase by the sputtering of the mantles because of the ions-neutral drift velocity \citep{gusdorf_sio_2008, jimenez-serra_parametrization_2008, lesaffre_low-velocity_2013,nguyen-luong_low-velocity_2013}.
Relevant for the case of the Fingers, the above models predict that shocks with velocities lower than about 10 km s$^{-1}$ are not strong enough to liberate Si or SiO from the frozen mantles  \citep[see, e.g.][]{nguyen-luong_low-velocity_2013}.

Nonetheless, it is worth emphasizing that these conclusions critically depend on the assumed frozen-SiO sputtering threshold energy, a parameter which is poorly constrained and which enters as an exponential factor in the equations \cite[e.g.][]{barlow1978,flower1994}.
For example, \cite{nguyen-luong_low-velocity_2013} used the  {Paris-Durham shock code}\footnote{The Paris-Durham shock code is publicly available at \url{https://ism.obspm.fr/shock.html}} where the SiO sputtering threshold energy is 1900 K \citep[in the original version of the code it was instead assumed to be 2500 K:][]{barlow1978,flower1994}.
We  {ran} the Paris-Durham shock code to simulate the conditions found in the Fingers, described above.
If the SiO sputtering threshold energy is  {1200 K }instead of 1900 K, SiO is predicted to be injected from the mantles into the gas-phase already at 7 km s$^{-1}$ with a predicted SiO column density consistent with what we observe in the Fingers.

However, it is important to emphasize that the above constraints have to be taken with a grain of salt and not at face value.
Therefore, giving the large uncertainty linked to the SiO sputtering threshold, which enters in an exponential term in the equations, one may speculate that shocks with velocities larger than about 7 km s$^{-1}$ can reproduce the observations.



\subsubsection{Shock velocity and inclination angle}
Using the time between Finger1 and Finger2 shock passage derived in Section \ref{subsec:time-constraint}, we can provide an approximation of the shock velocity:
\begin{equation}\label{eq:shock_vel}
\rm{v} \ \text{=} \   \dfrac{s_\perp}{\sin(\theta)}\dfrac{1}{time}
\end{equation}
where $s_\perp$ is the projected separation among the fingers and it is about 3000 au, and $\theta$ is the angle of the shock propagation with respect to the line of sight.

The lower limit to the shock velocity so to be able to release enough SiO into the gas-phase, reported in Section \ref{subsec:svel-constraint}, provides an upper limit to the inclination angle, which depends on the shock passage time.
Taking v$\geq 10$ km s$^{-1}$ gives  $\theta \leq$15, 25 and 45 degrees for 10$^4$, 5000 and 2000 yr, respectively.
In the case of lower SiO sputtering threshold, v$ \geq 7$ km s$^{-1}$ corresponds to  $\theta \leq$15, 25 and 90 degrees for 10$^4$, 5000 and 2000 yr, respectively.
Note that, if the time is 2000 yr, no solutions exist for a velocity lower then 7 \kms, for any $\theta$. 


\subsection{IRAS 4A fingers: the signature of the clash of an expanding bubble}\label{sub:discussion-origin}

As briefly mentioned in the Introduction, at a large scale, the NGC 1333 region is constituted by several filaments, of which the southeast one (SE) runs parallel to our Fingers \citep{dhabal_connecting_2019}. 
This SE filament is characterized by two substructures running parallel to each other with two distinct systemic velocities, +7.5 \kms \ and +8.2 \kms, respectively, with the western one blue-shifted with respect to the eastern \citep{dhabal_morphology_2018}. 
The gas in the arch-like structure west of the SE filament has approximately the same velocity as the western substructure of the SE filament, namely +6.5 \kms. 
This arch-like structure traces the borders of a large cavity (see Fig. \ref{fig:ch3oh+sio+hc3n_map}), to the north and west of which lie the protostellar systems of IRAS2, SVS13, and IRAS4 and to the south the SK1 system.
\citet{dhabal_connecting_2019} suggested that this region represents the gas compressed by a ``turbulent cell'' moving from the south and clashing against the NGC 1333 cloud. 
Always according to \citet{dhabal_connecting_2019}, this clash could have formed the SE filament and triggered the formation of the IRAS2, SVS13, and IRAS4 protostars.

The new detection of the train of shocks in the IRAS 4A region allows us to put some constraints to this hypothesis.

\begin{figure}
    \centering
    \includegraphics[scale=0.38]{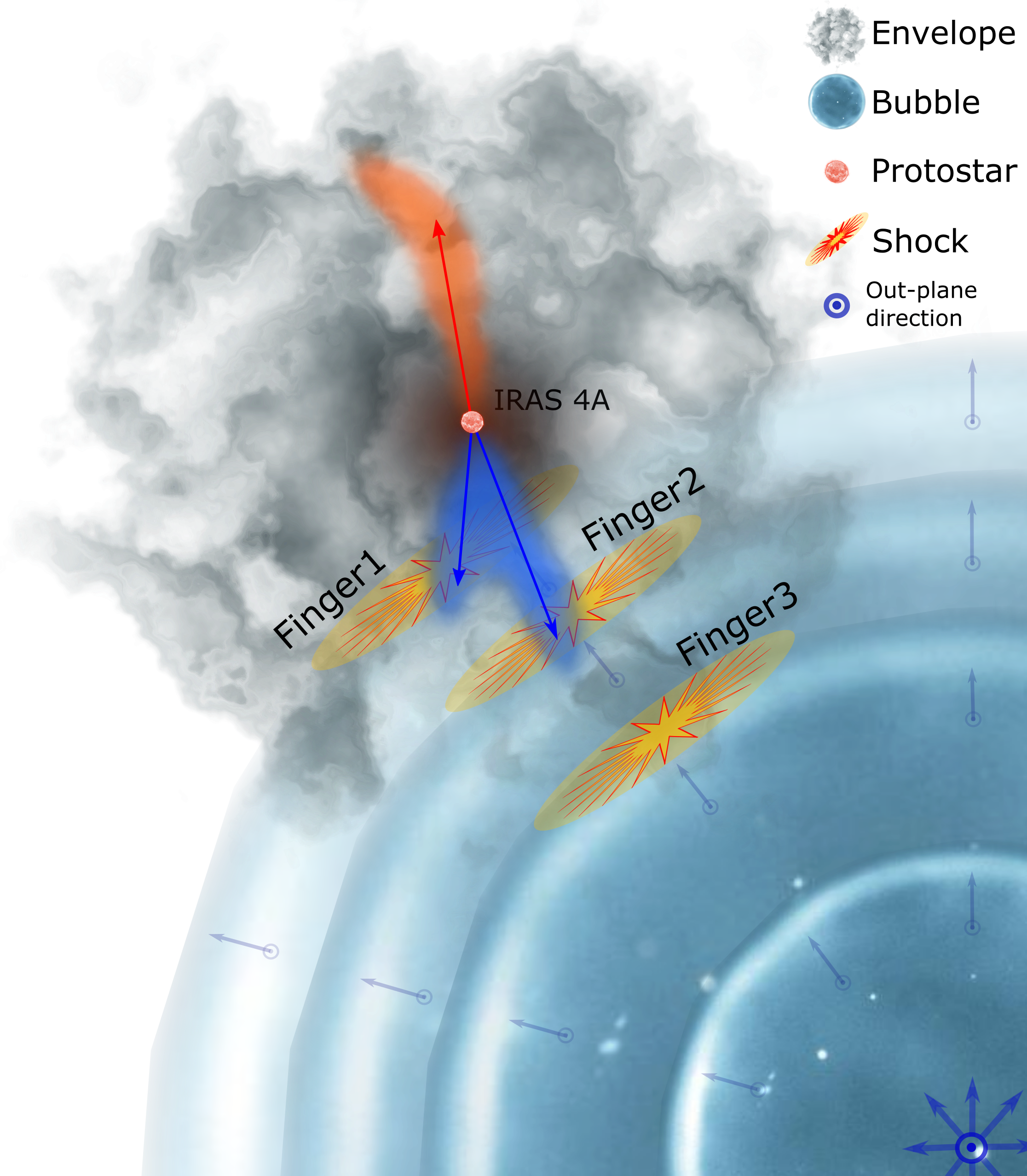}
    \caption{Cartoon model of the IRAS 4A region, illustrating the outflowing system inside the envelope and the proposed expanding bubble. 
    The  {size of the} bubble increases with time, and each circle in the figure represents the bubble at subsequent times: the largest circle represents the last expansion that caused the shock at the origin of Finger1. 
    In the cartoon, the bubble is farther from us in the line of sight than the IRAS 4A main cloud and it is moving toward it. 
    The three orange regions represent the  {collision zones}, in which the observed three fingers are formed.}
    \label{fig:cartoon_model}
\end{figure}

First, the presence of a train of three shocks in IRAS 4A suggests that the turbulent cell is the result of an expanding bubble, which caused three different shock events, as schematically shown in Fig. \ref{fig:cartoon_model}.
If the proposed bubble, coming from the southeast and moving toward the IRAS 4A main cloud from behind, continuously expands, it can create a series of shocks as the ones described in Sects. \ref{subsec:discussion-properties-summary} and \ref{subsec:discussion-train}.

Following the discussion done in Section \ref{subsec:discussion-train} to explain the observed SiO column density we propose that the expanding velocity of the bubble has to be at least 7 \kms.

Incidentally, this would not be the first case reported in the literature of a train of shocks caused by the clashing of an expanding bubble into a quiescent dense molecular cloud.
It has been observed, for example, in at least two other cases, both associated with supernovae remnants (SRN).
\citet{dumas_localized_2014} detected two parallel filamentary structures in SiO toward a cloud next to the SRN W51C and attributed them to the passage of the SNR primary shock.
Similarly, \citet{cosentino_interstellar_2019} reported the presence of two parallel shocks, probed by SiO, toward SNR W44, caused by the interaction of the expanding SNR bubble into the infrared dark cloud G034.77-00.55.

On the contrary, the origin of the expanding bubble clashing toward NGC 1333 is not obvious. 
As \citet{dhabal_connecting_2019} already pointed out, there are no known nearby ionizing stars, SNRs, or {\textsc{Hii}} regions that could be responsible for it.
Using CO isotopologues, \citet{arce_bubbling_2011} presented a complete mapping of shells and bubbles in the Perseus molecular cloud. 
Among them, southeast of IRAS 4A, there is the so-called CPS 2 shell with $\sim 6'$ of radius and whose origin is unknown.
However, CPS2 is located 24$'$ away from the IRAS 4A fingers, so it does not seem to be a good candidate.
On the other hand, our Galaxy is populated with very large (tens of degrees) loops, arcs, spurs, and filaments visible at different wavelengths (e.g., X-rays, microwaves, synchrotron emission).
The most famous are the so-called Loop I, II, III, and IV, probably old ($\sim 10^5-10^6$ yr) nearby supernova (SN) remnants \citep{berkhuijsen_are_1971, vidal_polarized_2015, dickinson_large-scale_2018}.
The NGC 1333 molecular complex lies at the edge of Loop II \citep[called also Cetus arc;][]{large_new_1962}, an expanding bubble of $\sim90^\circ$ of diameter and centered at galactic longitude $-100^\circ$ and latitude $30^\circ$. 
Hence, Loop II could be a possible candidate for the proposed expanding bubble. 

\subsection{Traces of other clashes in NGC 1333}\label{subsec:discussion-largescale}

\begin{figure*}
    \centering
    \includegraphics[scale=0.65]{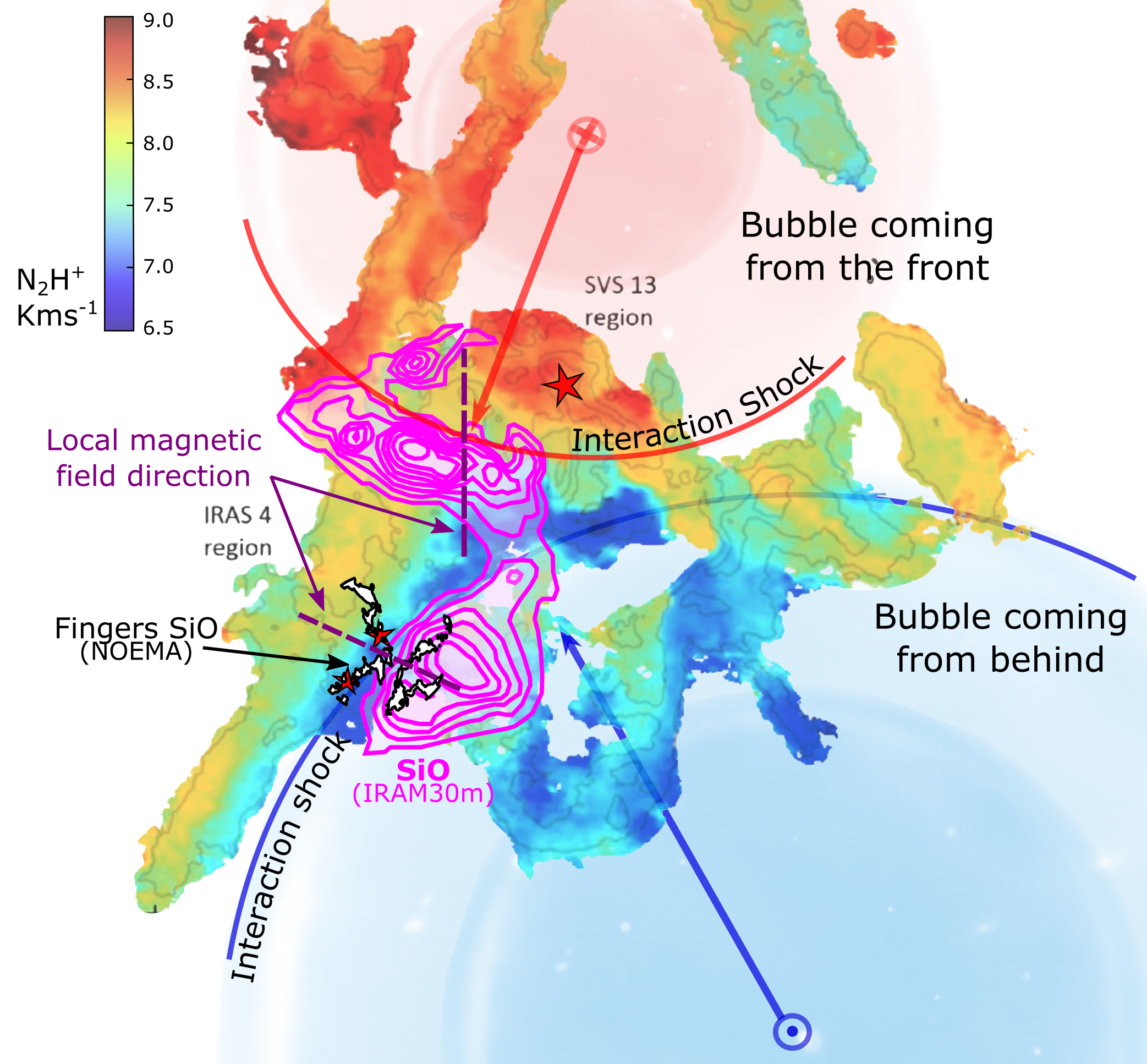}
    \caption{Large-scale view of the NGC 1333 filament. Overlap of the N$_2$H$^\text{+}$ line-of-sight velocity map by \citet{dhabal_connecting_2019} in color with the IRAM-30m SiO large scale ambient component emission by \citet{lefloch_widespread_1998} in magenta contours and the IRAS 4A fingers detected in SiO with the SOLIS/NOEMA observations of the present work in black contours. The purple dashed lines represent the local magnetic field direction from \citet{doi_jcmt_2020}. 
    The red stars mark the SVS13, IRAS 4A, and IRAS 4B protostars.
    The solid blue and red lines represent the interaction of the expanding bubbles: the red one is expanding from the northwest direction toward the SVS13 cloud from the front, and the blue one is expanding from the southwest direction toward the IRAS 4A cloud from behind.}
    \label{fig:dhabal+lefloch}
\end{figure*}
Previous single-dish observations have shown the presence of extended SiO emission, with a narrow (FWHM$<$1.5 \kms) line profile at ambient velocity, in the region encompassing SVS13 and IRAS 4A \citep{lefloch_widespread_1998,codella_low_1999}. 
The origin of this emission was debated but no clear consensus was reached.
For example, the hypothesis of fossil shocks connected with the outflows from the protostars was evoked, although no definite answer was found. 
The three fingers discovered by SOLIS can help to elucidate the origin of this large-scale SiO narrow emission.

Figure \ref{fig:dhabal+lefloch} shows the overlap of the line-of-sight N$_2$H$^\text{+}$ line-of-sight velocity map by \citet{dhabal_connecting_2019} of the southern NGC 1333 filaments with the large-scale SiO narrow emission by \citet{lefloch_widespread_1998} and the three fingers discovered by SOLIS in the present work.
The large-scale SiO narrow emission extends from southeast of SVS13A to southwest of IRAS 4A, and it seems to be composed of two parts, or lobes, that have different orientations.
In the northern lobe, the SiO emission is elongated in the direction northeast-west while, in the southern lobe, the emission runs in the direction northwest-southeast.
The SOLIS fingers lie at the northern limit of the southern lobe and they are pretty much parallel to the observed large-scale SiO emission.
As discussed in Sect. \ref{sub:discussion-origin}, we propose that the SOLIS fingers are due to shocks caused by the interaction of an expanding bubble coming from southeast behind the NGC 1333 SE filament, where IRAS 4A lies.
It is tempting to hypothesize that the southern lobe of the SiO emission observed by \citet{lefloch_widespread_1998} is an ensemble of small-scale shocks, like the ones traced by the SOLIS fingers, caused by older bubble expansion events.
Our prediction, therefore, is that the SiO emission of the southern lobe will break up in multiple parallel small-scale shocks if observed with interferometers like NOEMA.
If this is true, the frequency and chemical composition of these hypothetical shocks may add stringent constraints on the phenomenon and expansion velocity of the bubble clashing against NGC 1333 from the southeast.

A similar possibility could apply to the northern SiO lobe: it could be an ensemble of multiple parallel small-scale shocks unresolved by the single-dish observations of \citet{lefloch_widespread_1998} and \citet{codella_low_1999}.
However, in this case, since the SiO emission orientation is almost perpendicular to that of the southern lobe, the shocks would be caused by another bubble clashing from the northwest and expanding toward the region encompassing SVS13-A from the front (being that region red-shifted in the N$_2$H$^\text{+}$ velocity map). 
We searched for SiO and methanol fingers around SVS13-A using the SOLIS observations toward this source, but we could not find any.
Interestingly, however, the region encompassed by the SOLIS observations, in this case, lies in an area where \citet{lefloch_widespread_1998} and \citet{codella_low_1999} did not detect the SiO narrow emission, slightly north to where it appears.

Finally, the comparison of the large-scale SiO narrow emission map and the local magnetic field \citep[Figure 4 and 6 in ][]{doi_jcmt_2020} presents an intriguing feature (see Figure \ref{fig:dhabal+lefloch}).
The change in orientation observed in the two SiO lobes seems to be correlated with a change in the orientation of the magnetic field.
Specifically, in both cases, the magnetic field is almost parallel to the direction of the shocks, and almost perpendicular to the fingers/filamentary emission. 

\section{Conclusions}\label{sec:conclusions}

We report CH$_3$OH and SiO IRAM-NOEMA high spatial resolution ($\sim1\farcs5$; $\sim 450$ au) observations in the direction of NGC 1333 IRAS 4A, obtained in the context of the Large Program SOLIS.

The observations reveal the presence of three elongated filamentary structures traced by SiO and \meth \, $\sim 10''$ ($\sim$3000 au) south from the protostar center.
They are characterized by narrow (FWHM$\sim$1.5 \kms) lines peaked at the systemic velocity of the cloud.
These structures, which we called fingers, are parallel to each other, extended for more than about 6000 au, approximately equispaced by about 3000 au, and almost perpendicular to the two outflows arising from IRAS 4A and the SE filament where IRAS 4A lies.

The non-LTE analysis of the methanol lines in the northern finger indicates that the gas has a high density (5--20 $\times 10^5$ cm$^{-3}$) and high temperature (80--160 K), much larger than that expected if the gas was heated by the central protostar. 
The three detected fingers are chemically different, with the northern one traced by both SiO and \meth \ and the southern two only by SiO.
The CH$_3$OH over SiO abundance ratio is 160--300 in the northern finger, while it is $\leq 40$ in the southern one.
Both the measured temperature and enhanced SiO and \meth \ abundances point to a non-thermal process responsible for them.

Given their quasi-periodicity and morphology, we considered the possibility that the fingers trace a Kelvin-Helmholtz instability occurring at the interface of the NGC 1333 IRAS4 region and a less dense cloud sliding from south to north.
However, this hypothesis can not reproduce the observed properties of the fingers.

We then considered the possibility that the three fingers represent a train of three shocks. 
This hypothesis agrees with the observed physical and chemical properties of the fingers, and provides constraints on when the shocks occurred, with an interval of at least 5000 yr between the youngest northern finger and the next southern one.

Previous studies had already shown that the NGC 1333 is a region heavily shaped by the dynamical interaction of internal outflows and external bubbles with the quiescent molecular cloud.
In particular, previous large-scale maps of gas distribution and velocities in the IRAS 4A region already suggested the presence of a ``turbulent cell'' pushing toward the NGC 1333 IRAS4 from behind and coming from the south.
The newly detected fingers provide support to this hypothesis, considering that the turbulent cell is the result of an expanding bubble, which caused three different shock events, with an expanding velocity of at least 7 \kms.

Finally, we suggest that the widespread narrow SiO emission observed toward the NGC 1333 IRAS 4 and SVS 13 region with single-dish observations in the late 90s is due to unresolved trains of shocks like the SOLIS fingers.
These shocks would be the signature of the interaction of the bubble giving rise to the IRAS 4A fingers in the south and of another bubble pushing from the north toward SVS 13.
We predict, therefore, that large-scale high-spatial resolution mosaics of SiO and other shock-related species can help to fully reconstruct the dynamical history of NGC 1333 and, consequently, provide precious constraints to the relevant theories and models.

\section*{Acknowledgments}
 {We thank the referee P. Goldsmith for the fruitful comments and suggestions.}
We are very grateful to all the IRAM staff, whose dedication allowed us to carry out the SOLIS project.
We warmly acknowledge fruitful discussion with Prof. Lucio Piccirillo and Dr. Geoffroy Lesur, as well as with Dr. Antoine Gusdorf on the shock models.
Some of the computations presented in this paper were performed using the GRICAD infrastructure (\url{https://gricad.univ-grenoble-alpes.fr}), which is partly supported by the Equip@Meso project (reference ANR-10-EQPX-29-01) of the programme Investissements d'Avenir supervised by the Agence Nationale pour la Recherche.
This project has received funding within the European Union’s Horizon 2020 research and innovation programme from the European Research Council (ERC) for the project “The Dawn of Organic Chemistry” (DOC), grant agreement No 741002, and from the Marie Sklodowska-Curie for the project ``Astro-Chemical Origin'' (ACO), grant agreement No 811312.
C. Codella acknowledges the project PRIN-INAF 2016 The Cradle of Life - GENESIS-SKA (General Conditions in Early Planetary Systems for the rise of life with SKA).

\section*{Data Availability} 
The data underlying this article are part of the IRAM-NOEMA SOLIS Large program and they will be publicly available on the IRAM archive at the end of the proprietary period (\url{https://www.iram-institute.org/EN/content-page-240-7-158-240-0-0.html}).


\bibliographystyle{mnras}
\bibliography{Narrows_Outflow}





\bsp	
\label{lastpage}
\end{document}